\documentclass[10pt]{iopart}
\usepackage{iopams}
\usepackage{graphicx}
\usepackage[english]{babel}

\makeatletter

\begin{document}

\topical{X-ray imaging of spin currents and magnetisation dynamics at the nanoscale}

\author{Stefano Bonetti$^*$}
\address{$^*$ Department of Physics, Stockholm University, Sweden}
\ead{stefano.bonetti@fysik.su.se}

\begin{abstract}
Understanding how spins move in time and space is the aim of both fundamental and applied research in modern magnetism. Over the past three decades, research in this field has led to technological advances that have had a major impact on our society, while improving the understanding of the fundamentals of spin physics. However, important questions still remain unanswered, because it is experimentally challenging to directly observe spins and their motion with a combined high spatial and temporal resolution.
In this article, we present an overview of the recent advances in X-ray microscopy that allow researchers to directly watch spins move in time and space at the microscopically relevant scales. We discuss scanning X-ray transmission microscopy (STXM) at resonant soft X-ray edges, which is available at most modern synchrotron light sources. This technique measures magnetic contrast through the X-ray magnetic circular dichroism (XMCD) effect at the resonant absorption edges, while focusing the X-ray radiation at the nanometre scale, and using the intrinsic pulsed structure of synchrotron-generated X-rays to create time-resolved images of magnetism at the nanoscale. In particular, we discuss how the presence of spin currents can be detected by imaging spin accumulation, and how the magnetisation dynamics in thin ferromagnetic films can be directly imaged. We discuss how a direct look at the phenomena allows for a deeper understanding of the the physics at play, that is not accessible to other, more indirect techniques. Finally, we present an overview of the exciting opportunities that lie ahead to further understand the fundamentals of novel spin physics, opportunities offered by the appearance of diffraction limited storage rings and free electron lasers.
\end{abstract}
%\pacs{}
%\keywords{X-rays, magnetism, spin currents, spin waves}
\submitto{\JPCM}

\maketitle
\ioptwocol

\section{Introduction}
A large part of research in modern magnetism deals with \emph{nanomagnetism}, i.e. with those magnetic phenomena that take place at the nanometre scale. Nanomagnetism is a fascinating field of investigation that has advanced rapidly in the last three decades thanks to an almost unique interplay between fundamental and applied research. Such an interplay has led to several breakthroughs that have greatly influenced our understanding of the fundamentals of magnetism and, in a few cases, even impacted society at large. Following the discovery of the giant magnetoresistance effect \cite{baibich1988giant,binasch1989enhanced} and its implementation in all hard disk drives \cite{parkin1990oscillations,parkin1991oscillatory,dieny1991magnetotransport,parkin1993origin,tsang1994design}, the end of the millennium saw the fundamental discovery of another phenomenon that can occur in nanometre-sized magnetic objects: the spin transfer torque effect. This effect was predicted by Slonczewski and Berger in 1996 \cite{slonczewski1996current, berger1996emission} and experimentally observed in 1999 \cite{myers1999current}. Just a few years ago, the first device based on the spin transfer torque was commercialised by Everspin Technologies \cite{slaughter2012high}, and other companies are planning to enter the market \cite{wolf2014time,nowakvoltage}. In addition, there have been several other key fundamental discoveries that are currently challenging our understanding of magnetism, and that have the potential for applications in the future: \emph{i)} the observation of ultrafast (sub-ps) magnetisation dynamics driven by femtosecond laser pulses \cite{beaurepaire1996ultrafast} and all-optical switching of magnetisation \cite{stanciu2007all}; \emph{ii)} the experimental realization of the spin Hall effect in metals \cite{valenzuela2006direct,saitoh2006conversion}; \emph{iii)} the discovery of topological insulators \cite{zhang2009topological}. In all cases, there is an opportunity for the implementation of denser, faster and more energy efficient information storage and manipulation.

One of the key concepts in the field of nanomagnetism is \emph{spin current}, which is the notion that spin angular momentum can flow in an ordered fashion and be used to manipulate magnetic order without the need for magnetic fields, which are generally energy-hungry and difficult to implement at the nanoscale. Great progress has been made in understanding how a spin current can be generated, transported and detected by electrical means, and, ultimately, used to manipulate the magnetisation in nanomagnets. However, it is still challenging to directly image the phenomena resulting from the flow of a spin current in nanostructures comprising magnetic layers, in particular the appearance of spin accumulation and the occurrence of magnetisation dynamics. One must use a tool that has not only magnetic sensitivity, necessary to see spins, but that also combines both high spatial and temporal resolution. There are a few techniques that possess two of these three characteristics, such as femtosecond magneto-optics \cite{kirilyuk2010ultrafast} (magnetic sensitivity, fast) and magnetic force microscopy \cite{koblischka2003recent} or electron microscopy \cite{doi:10.1063/1.2799420} (magnetic sensitivity, high spatial resolution), but the combination of all three is still challenging. Time-resolved X-ray microscopy at synchrotron light sources combines all these three characteristics at once. Synchrotron-generated X-rays have: a wavelength on the order of 1 nm, which allows for focusing to the nanometre scale; a variable polarisation that enables magnetic contrast through the X-ray magnetic circular dichroism (XMCD) effect at the resonant absorption edges; and an intrinsic pulsed structure that can provide the time resolution.

The goal of this Topical Review is to present the progress made in scanning transmission X-ray microscopy (STXM) at synchrotron light sources and to give an outlook over the opportunities ahead. It is a particularly exciting moment for the research community since the first diffraction limited storage ring (DLSR) started its operation \cite{eriksson2014diffraction}, greatly enhancing the brightness of synchrotron X-rays, and since free electron lasers (FELs) have been able to produce fully coherent photons in the soft and hard X-ray regime at the femtosecond time scales \cite{emma2010first,ishikawa2012compact}. Thanks to these new experimental tools, the future of time-resolved X-ray imaging looks bright. It is reasonable to expect that researchers will be able to make ``movies'' of how spins move in time and space with ever increasing temporal and spatial resolutions, and sensitivity, which in turn is expected to lead towards a deeper understanding of the fundamentals of magnetism.

The structure of this Topical Review is as follows. The fundamentals of spin currents, magnetisation dynamics and X-rays are revised in Section 2. In Section 3, we describe the working principle of a STXM setup, currently available at most synchrotron light sources, that allows for spin-sensitive imaging with time-resolution. In section 4, we review the experimental efforts in imaging spin currents and spin-current induced switching of magnetisation, while in Section 5, we review experiments related to the imaging of magnetisation dynamics. An outlook of the future challenges in the field of spin physics, as well as the new opportunities offered by diffraction limited storage rings and free electron lasers are discussed in Section 6.

\section{Spin currents, magnetisation dynamics and X-rays: Key concepts}

The goal of this section is to review the important theoretical concepts needed to understand the imaging experiments presented in Sections 4 and 5. In the first two subsections, we review the fundamentals of the physics of spin transport and magnetisation dynamics. Key notions within the fields of magnetism and magneto-transport, such as magnetisation, spin current and the Stoner model, are assumed to be familiar to the reader. It is also expected that the concepts of spin accumulation, spin transfer torque and the Landau--Lifshitz--Gilbert (LLG) equation have been encountered before. In the last subsection, we briefly present the theory necessary to understand how X-rays can ``see'' spins. Readers interested in a detailed understanding of the interaction between polarised X-rays and matter are directed towards a more comprehensive text, such as Ref.~\cite{stohr2007magnetism}.

\subsection{Spin currents: From magnetoresistance to spin transfer torque}
The concept of spin current is key to modern magnetism. The simplest way of picturing a spin current is to imagine a flow of electrons in a conductor with all their spins pointing in a well-defined direction. In this case, the spin current, which is the flow of net spin angular momentum, is collinear and proportional to the charge current. In other words, each electron carries its $\hbar/2$ of angular momentum along with it. Indeed, this is the situation considered throughout most of this Topical Review. However, as we mention in Section 6, this does not necessarily have to be the case; the flow of charge and spin can be decoupled, as for the case of the spin Hall effect.

\begin{figure}[t]
%\begin{figure}[p]
\centering
\includegraphics[width=0.8\columnwidth]{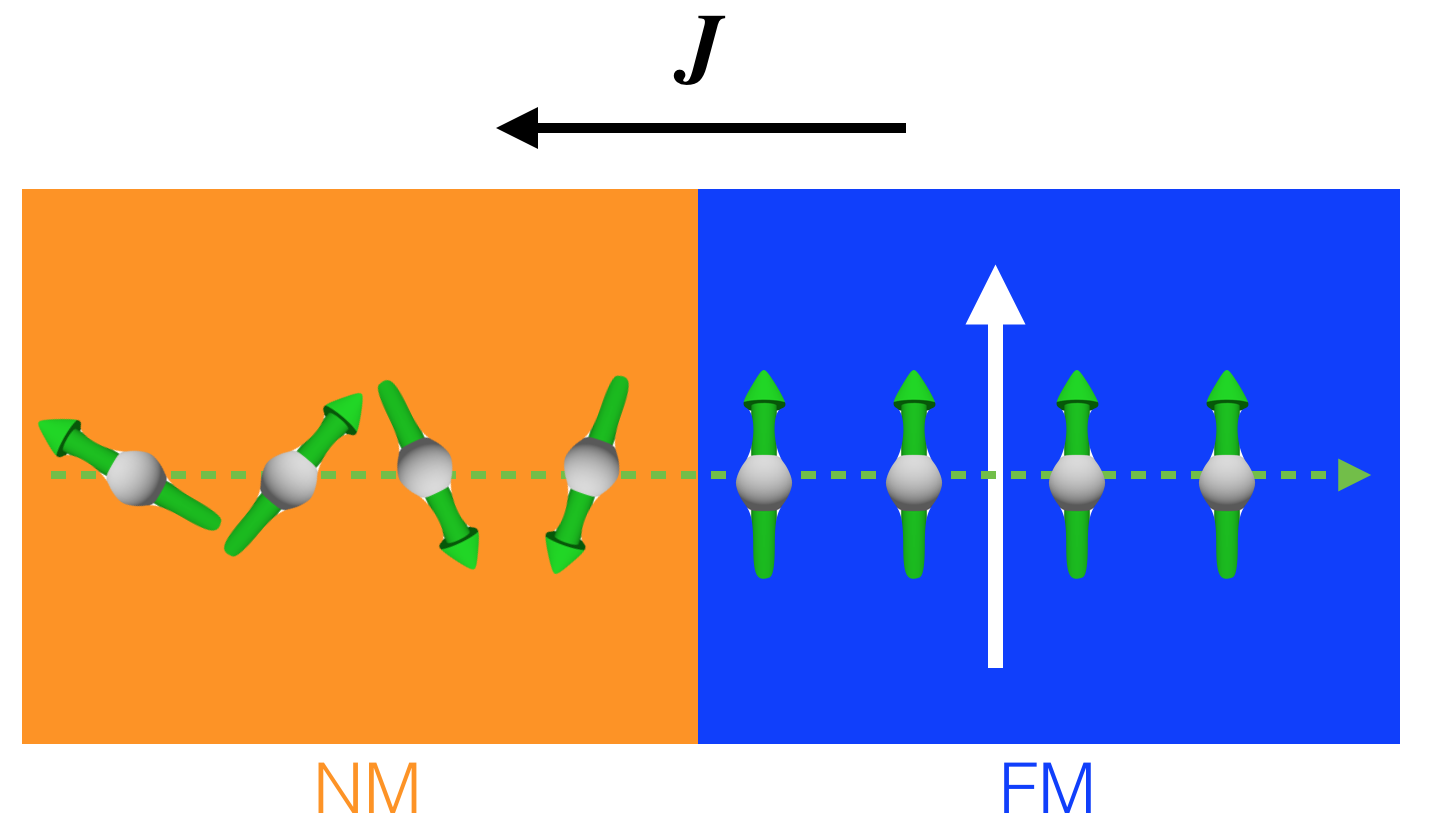}
\caption{Schematic of the current flow in a non-magnetic (NM) metal and in a ferromagnetic (FM) one. The current in the non-magnetic metal has no net spin polarisation, while the flow of charge in a ferromagnet is associated with a flow of spin angular momentum. The spin polarisation is collinear to the magnetisation of the ferromagnet, indicated by the white arrow. Note that the current density $J$ and the flow of electrons have opposite signs.}
\label{fig:spinpol}
\end{figure}

The most practical way to create a spin current is to let a spin-unpolarised (or simply ``unpolarised'') charge current flow through a ferromagnetic metal. The electrons exiting the ferromagnet have their spins collinear with the magnetisation of the ferromagnet and, in turn, also carry a net flow of angular momentum, as shown schematically in Fig.~\ref{fig:spinpol}. We stress the use of the word \emph{collinear} rather than parallel or antiparallel, because the sign of the spin polarisation in transport experiments depends on the density of states at the Fermi level (qualitatively different for Fe as compared to Co and Ni for instance) and on the $s$$d$ scattering. The sign of the spin polarisation can be measured experimentally using spin-polarised photoemission  \cite{busch1969observation,banninger1970photoelectron} or transport \cite{vouille1999microscopic} experiments. Theoretically, it can be determined only by performing accurate calculations on the specific system of interest. Such considerations are well beyond the scope of this article, and we direct the interested reader towards a more comprehensive discussion on the topic, such as Ref. \cite{stohr2007magnetism}. Since we aim to provide a qualitative description of the phenomenon, we will keep it simple and draw the majority spins and magnetisation parallel to each other with the understanding that in some systems, the situation could be the opposite.\footnote{There is another important note on the convention used for majority and minority spins. For visual clarity, majority spin states and magnetisation in a ferromagnet are often drawn parallel to each other. However, the magnetisation is defined as $\mathbf{M}=\mathbf{m}/V$, where $\mathbf{m}$ is the magnetic moment and $V$ is the volume of the sample. The magnetic moment is defined as $\mathbf{m}=-g\mu_B\mathbf{s}$, which means that spins and magnetisation point in opposite directions, or that the magnetisation is aligned with the \emph{minority} spins. This discussion is left out of the main text to avoid another sign flip to keep track of. None of the important qualitative discussions presented here depend on such sign.}

The spin-dependent scattering of the conduction electrons is the key mechanism that creates spin polarisation, which can be understood in terms of the Stoner model \cite{stohr2007magnetism}. The band structure of a $3d$ metallic ferromagnet such as Fe, Co and Ni, can be described in terms of an ``itinerant'', hybridized $sp$ band and a ``localised'' $d$ band. The Stoner model assumes that the $d$ band is split into two spin $d$ bands shifted in energy ($\sim1$ eV) due to the exchange interaction. Since the Fermi energy in a material is fixed, the material contains more spins of the band that is shifted lower in energy, which are named \emph{majority} spin states. Conversely, the spins in the band shifted higher in energy are called \emph{minority} spin states. Since the total number of states in a given band is fixed (10 states/atom for $d$ bands), there are more empty states (i.e. holes) available in the minority band. What this means is that the itinerant minority electrons in the $sp$ band have a higher probability to scatter in an empty $d$ state than the majority electrons, since electron-electron scattering events are spin-conserving and, thus, spin-flip events are forbidden. In other words, the flow of unpolarised electrons into the ferromagnet results in the minority spin component experiencing a greater resistance in moving through the material, while the majority spins can flow less perturbed. The net effect is that, at the exit of the ferromagnet, the original unpolarised current (i.e. made up of an equal number of majority and minority spins) is spin polarised. The quantization axis for the spins is given by the direction of the magnetisation $\mathbf{M}$ in the material, which can be set by an external magnetic field. Another way of looking at this, according to Mott \cite{mott1936electrical}, is that the conductivity in a ferromagnet can be modeled as an electric circuit made up of two resistors connected in parallel, one for each spin channel. The total conductivity is then computed as for any electric circuit as the sum of the inverse resistivity of the two spin channels.

So far, we have discussed electrons entering and exiting a ferromagnet in an abstract fashion. In practice, this is often achieved by placing a thin film ferromagnet in contact with non-magnetic conductors, such as Cu, and use these conducting layers to inject and detect electrons. The technology for depositing layers of thin metallic films is so well developed that well-defined interfaces can be assumed in the description of most phenomena.

An unpolarised current in a non-magnetic metal (NM) can be thought to be made up of an equal amount of majority and minority spins. When an unpolarised current flows from a non-magnetic material into a ferromagnet, the majority and minority spins have different scattering probabilities. Spins that are parallel to the spin polarisation in the ferromagnet have a higher probability of being transmitted through the ferromagnet, while antiparallel spins have a higher probability of being reflected from the interface. This is the so-called \emph{spin filter effect}. This creates an apparently contradictory situation. Far away from the interface, in the non-magnetic metal, there must be an equal number of parallel and antiparallel spins with respect to the spin polarisation in the ferromagnet. However, the net flow of reflected antiparallel spins seems to imply their buildup in the non-magnetic metal. In reality, at relatively long distances from the interface ($\sim100$ nm for Cu), spins diffuse and electrons become effectively unpolarised. However, closer to the interface, there is an \emph{accumulation} of antiparallel spins, which is only caused by the spin-dependent transmissivity and reflectivity of the interface. This effect is called \emph{spin accumulation}, and the length over which it decays is called the \emph{spin diffusion length}. Naturally, spin accumulation also occurs when electrons leave a ferromagnet and are injected into a non-magnet. In this case, the spins parallel to the spin polarisation accumulate at the interface with the non-magnetic metal over the same spin diffusion length. In many interfaces of interest, such as Cu/Co or Cu/Ni \cite{stiles2002anatomy}, the spin polarisation in the ferromagnet is parallel to the majority spins.

We have thus far neglected the conservation of angular momentum in the discussion. However, when creating a spin-polarised current, the spin angular momentum cannot be created out of nowhere. The source of angular momentum is, of course, the ferromagnet itself, which implies that, for the angular momentum to be conserved, the ferromagnet has to depolarise. This is correct, although it is important to consider the relevant limits.

\begin{figure}[t]
%\begin{figure}[p]
\centering
\includegraphics[width=\columnwidth]{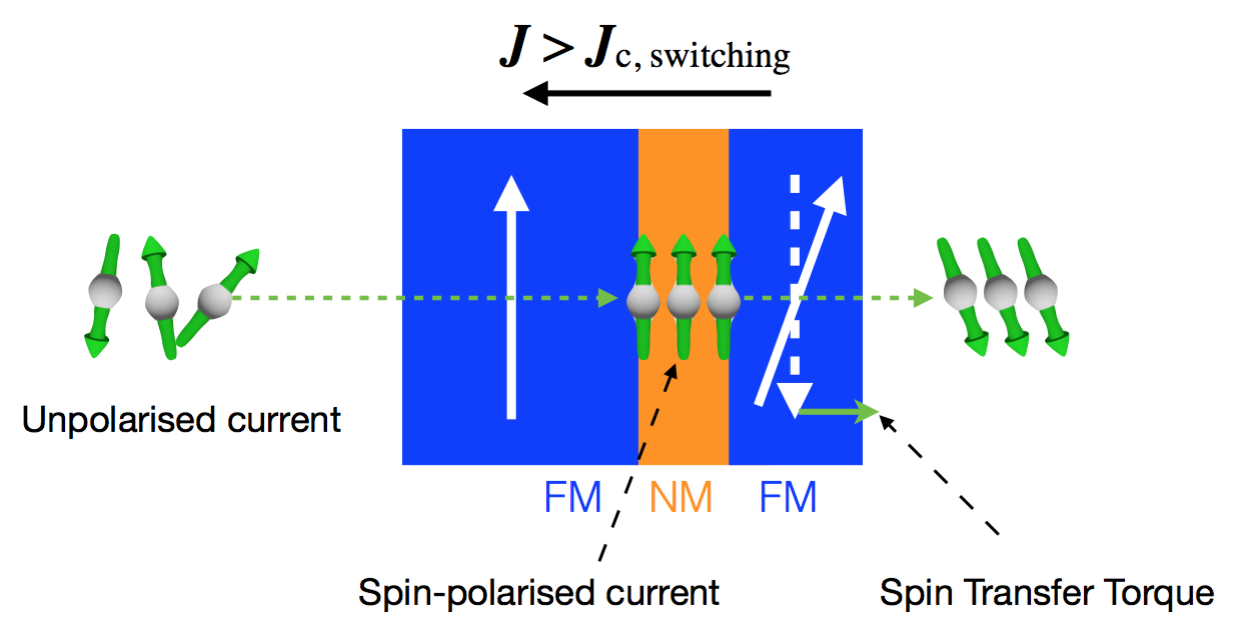}
\caption{Schematic of a spin-transfer torque induced switching mechanism in an asymmetric spin valve with lateral dimensions of the order of 100 nm. The thicker ferromagnet acts as a polariser and a ``fixed'' reference layer, while the thinner ferromagnetic layer is ``free'' to rotate under the effect of a torque. A large unpolarised current density $\mathbf{J}$ greater than a critical value $\mathbf{J}_{\textrm{c, switching}}$ is sent through the spin valve, where the two layers have antiparallel magnetisation. The ``fixed'' layer polarises the electrons, which can exert a net torque on the ``free'' magnetic layer and align it parallel to the ``fixed'' layer. Reversing the current causes the switching from the parallel to the antiparallel configuration to occur.}
\label{fig:STT_switch}
\end{figure}

For current densities below $10^6$ A/cm$^2$, the depolarisation of the ferromagnet is negligible, and it takes place at a slow enough rate that the Gilbert damping in the ferromagnet manages to compensate it. The net effect is that the current becomes spin-polarised with no significant change in the ferromagnet. The situation is more interesting when the current density is greater than $10^6$ A/cm$^2$ and it is spin-polarised. When this net flow of spin angular momentum is sent through a ferromagnet, it can exert a significant torque on the magnetisation, which is called the \emph{spin transfer torque} effect, as predicted by Slonczewski and Berger two decades ago \cite{slonczewski1996current, berger1996emission}.\footnote{One possible and interesting way to estimate the current density needed to observe a measurable torque is as follows. For the spin-polarised current to affect the magnetisation in the ferromagnet, the number of the flowing spins has to be roughly the same number as the spins in the ferromagnet. However, these spins cannot take an arbitrarily large amount of time to flow; they have to exert a torque on the magnetisation before the Gilbert damping torque restores the equilibrium condition. Assuming a ferromagnet with 1 $\mu_B$/atom and a density $\sim$$10^{23}$ atoms/cm$^3$, and a Gilbert damping rate $\Gamma=\alpha\nu_0\approx100$ MHz (reasonable for a Gilbert damping $\alpha\approx0.01$ and a ferromagnetic resonance frequency $\nu_0\approx10$ GHz), the net flow of spins into the volume of the ferromagnet needs to be at least $10^{23}$ spins/cm$^3$ every 10 ns (1/100 MHz). Assuming a perfectly polarised spin current (i.e. 1 $\mu_B$/electron), this can be written as $10^{12}$ A/cm$^3$. Since we are dealing with thin films with thicknesses $\sim$$10$ nm or $10^{-6}$ cm, this corresponds to a current density of $10^6$ A/cm$^2$. This assumes perfectly effective spin transfer torque; the first devices operated with $10^7$-$10^8$ A/cm$^2$. However, in recent commercial devices thinner ($\sim1$ nm) layers and optimised structures have been implemented, current densities of $10^6$ A/cm$^2$ are sufficient to observe spin torque switching.}  In his seminal paper \cite{slonczewski1996current}, Slonczewski predicted that the spin transfer torque could be large enough to be used to switch the magnetisation of a nanosized ferromagnet without the need of magnetic fields. The schematic of a device based on spin transfer torque is illustrated in Fig.~\ref{fig:STT_switch}.

In order for the spin transfer torque to be dominant over other current-induced effects, such as heating and Oersted fields, the current needs to flow in small regions. Typically, nanostructures of $\sim100$ nm in lateral size or smaller are necessary, in which currents in the mA range are sufficient to achieve the required current densities, while limiting the heat deposited in the sample.

\subsection{Magnetisation dynamics: From ferromagnetic resonance to current-induced spin waves}
Ferromagnetic resonance (FMR) is the uniform precession of spins that can be observed in ferromagnets subjected to oscillating magnetic fields, typically in the GHz range. The resonance is present because the magnetization obey the LLG equation of motion:
\begin{equation}
\frac{d\mathbf{m}}{dt}=-\gamma\mathbf{m}\times\mathbf{H}_{\rm eff}+\alpha\frac{d\mathbf{m}}{dt}\times\mathbf{m},
\label{eq:llg}
\end{equation}
where $\gamma=g\hbar/\mu_B\approx28$ GHz/T is the gyromagnetic ratio and $\alpha$ is the Gilbert damping. The resonance frequency depends on the magnitude and direction of the magnetisation and the applied (static) magnetic field,  the shape of the ferromagnet and its internal anisotropies. Typically, one includes all these different parameters to calculate the total energy in the ferromagnet and then computes the effective magnetic field $\mathbf{H_{\rm eff}}={\delta E}/{\delta M}$.

Measurements of the FMR are extremely useful for the characterisation of thin magnetic films. In particular, key parameters, such as the saturation magnetisation $M_s$ and the Gilbert damping $\alpha$ can be retrieved in a relatively straightforward way. Conventional techniques measure FMR by exciting the magnetisation of a ferromagnet with an AC magnetic field and measuring the electromagnetic power absorbed. The resonance condition is found either by sweeping the frequency of the AC magnetic field and keeping the larger DC-applied magnetic field constant, or by sweeping the magnitude of the DC field and keeping the frequency constant.

As the amplitude of the driving AC magnetic field is increased, however, the response of the magnetisation saturates well below the geometrical limit. This is due to the occurrence of the so-called Suhl instabilities \cite{suhl1957theory}, or the coupling of the uniform FMR mode to the non-uniform modes, which are degenerate in frequency \cite{silva2008developments}. In spin transfer torque experiments, the nanoscale confinement of the excitation, which is necessary to achieve high enough current densities, largely modifies the spin wave spectrum, allowing for the uniform precession mode to reach and overcome geometrical saturation \cite{bonetti2012power}.

\begin{figure}[t]
%\begin{figure}[p]
\centering
\includegraphics[width=\columnwidth]{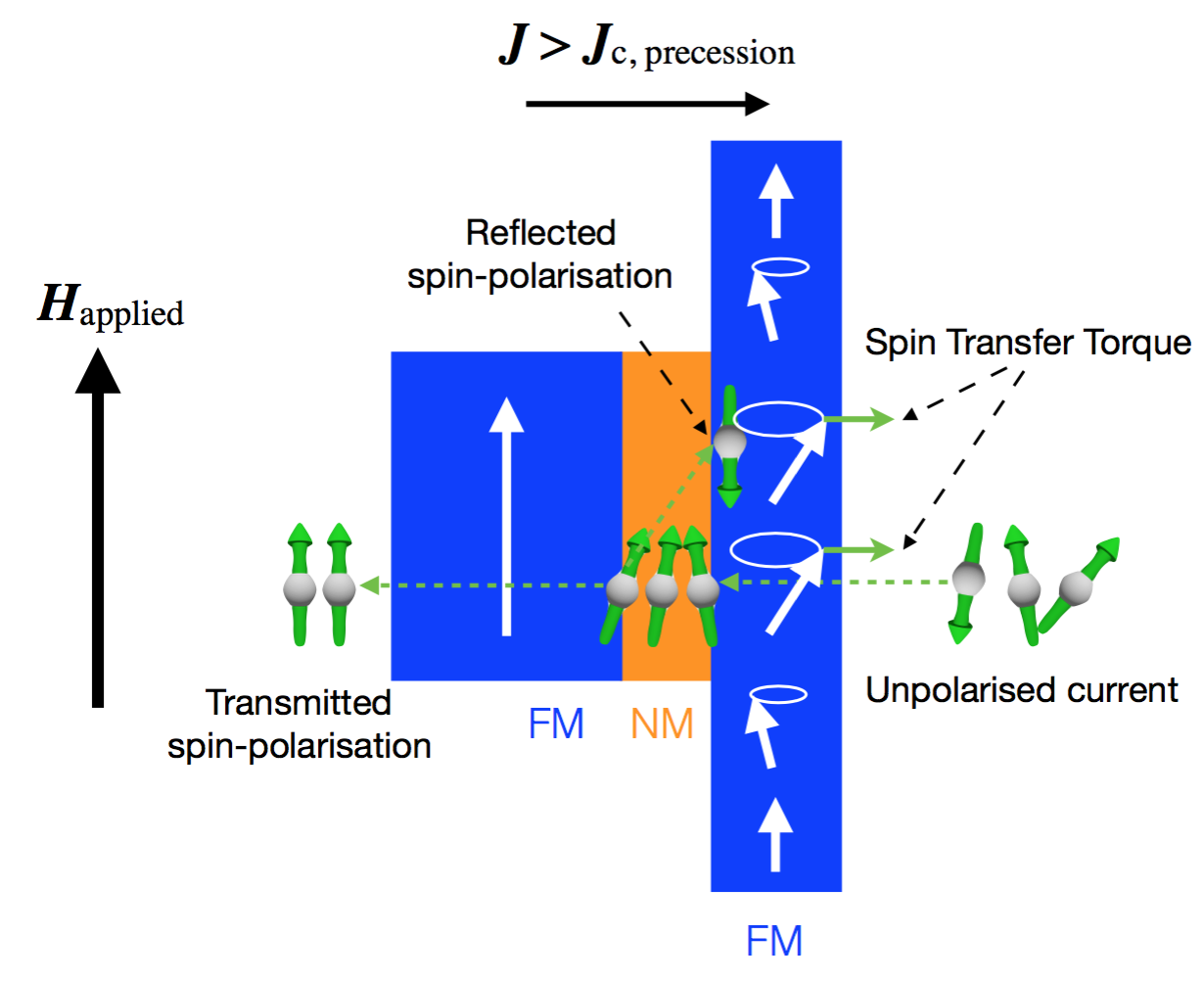}
\caption{Schematic of the excitation of spin waves due to the spin transfer torque in an asymmetric spin valve. Here, only the ``fixed'' layer is patterned to a lateral dimension $\sim100$ nm, while the ``free'' layer is extended over micrometre lengths. A large unpolarised current density $\mathbf{J}$ greater than a critical value $\mathbf{J}_{\textrm{c, precession}}$ is sent through the structure (with the electrons flowing from right to left), where the two layers have almost parallel magnetisation. The thin ``free'' layer only partly polarises the electrons flowing through it, which are then filtered by the ``fixed'' reference layer. The transmitted electrons have spin ``up'' polarisation, while the electron \emph{reflected} at the FM/NM interface have spin ``down'' polarisation. It is these spin ``down'' electrons that flow towards the ``free'' layer and tend to destabilise the magnetisation. With the proper combination of applied magnetic field and current, a magnetisation precession can be sustained.}
\label{fig:STT_sw}
\end{figure} 

Slonczewski showed that Eq.~(\ref{eq:llg}) can be rewritten to include an extra term:
\begin{eqnarray}
\frac{d\mathbf{m}}{dt}&=&-\gamma\mathbf{m}\times\mathbf{H}_{\rm eff}+\alpha\frac{d\mathbf{m}}{dt}\times\mathbf{m} \nonumber\\
 &+& \eta(\theta)\frac{\mu_B}{e}\frac{J}{t}\mathbf{m} \times\left(\mathbf{m}\times\mathbf{M}\right),
\label{eq:llgs}
\end{eqnarray}
where $J$ is the current density, $t$ is the film thickness, $\eta(\theta)$ measure the efficiency of the spin torque transfer, and $\mathbf{M}$ is the normalised magnetisation of the ``fixed'' layer that defines the spin-polarisation of the current. Without going into much detail of Eq.~(\ref{eq:llgs}), referred to as the Landau--Lifshitz--Gilbert--Slonczewski (LLG + S) equation, one can identify a few of its qualitative features. First, the new term vanishes when the magnetisation of ``free'' and ``fixed'' layer are collinear, and it is maximised when the magnetisations are orthogonal to each other. Second, the Slonczewski term comprises a current density term, which means that its sign can be reversed by reversing the current polarity. Third, it can be shown that, in the appropriate limit and with the current flowing with the correct polarity, the Slonczewski term can cancel out the Gilbert damping term. The net effect is that the right-hand side of Eq.~(\ref{eq:llgs}) reduces to its first term, i.e. the magnetisation keeps precessing around the effective field as long as the current is applied. This cancellation is not achieved exactly at every instant of time, but over one precession period on average. In terms of nonlinear dynamics, the solution of the LLG + S equation is a limit cycle \cite{bertotti2005magnetization}. For completeness, it should be noted that in magnetic tunnel junctions, spin transfer torque can also give rise to substantial field-like terms in the LLG + S equation \cite{zhang2002mechanisms}, or terms that have the form of an effective additional magnetic field around which the magnetisation precesses. However, these terms are typically one or more orders of magnitude smaller than the Slonczewski term (current-like) in metallic mutilayers, which are the ones described in this review.

A schematic of the mechanism of excitation of magnetisation dynamics due to spin transfer torque is presented in Fig.~\ref{fig:STT_sw}. All the qualitatively important features have been included in the diagram and are described in the caption. When the ``free'' layer is extended outside the region where the current is focused, typically by patterning the lead or the ``fixed'' layer to the nanometre scale, the excited magnetisation dynamics is non-uniform; the precession amplitude is largest below the region of current flow, and it gradually decreases as the distance from that region increases. Two types of spin torque dynamics are typically observed: \emph{i)} vortex oscillations, typically present in bigger nanocontacts where the relative effect of the Oersted field is larger; \emph{ii)} spin waves with either propagating or localised character, mostly excited in smaller structures where the spin transfer torque dominates.

\subsection{How can one ``see" spins with X-rays? The fundamental role of spin-orbit coupling for the X-ray magnetic circular dichroism (XMCD) effect.}
Before we discuss the details of X-ray microscopy, it is worthwhile to reflect on how it is possible to see something magnetic using photons. Recalling concepts from basic electromagnetism, at low frequencies, both the electric and the magnetic part of the electromagnetic radiation can interact with matter, depending on the magnitude of the magnetic permeability and of the dielectric permittivity. For ferromagnets, the magnetic permeability can be extremely large at low frequencies; hower, in the $100$ GHz -- 1 THz region, the permeability is 1 for all materials. The typical pictorial explanation offered is that, as the frequency of the electromagnetic field increases, it is progressively harder for spins to follow the oscillation of the magnetic field component. In a more quantitative way, this relates to the almost fixed magnitude of the gyromagnetic ratio $\gamma=g\mu_B/\hbar\approx28$ GHz/T for electrons, where $g$ is the Lande g-factor, $\mu_B$ is the Bohr magneton and $\hbar$ is the Planck constant, i.e. that is determined only by fundamental constants. The fact that there are no magnetic monopoles (i.e. the magnetic moment of an electron is quantized in units of $\mu_B$), and that the spin of the electron has a well-defined angular momentum $\hbar/2$ that needs to be conserved, is what suppresses the amplitude of the spin precession at high frequencies. X-rays have frequencies of $10^{5}$--$10^{6}$ THz ($1$--$10$ keV), and we can therefore safely exclude a significant interaction between the spins and the magnetic component of the X-ray radiation.

Therefore, the electric component of the X-ray radiation must be the one that ``sees'' the spin, although there are no explicit spin-dependent terms in an electric dipole operator. This seems to be a paradox. The solution to the paradox is found by taking into consideration the spin-orbit coupling, the relativistic interaction that couples the spin of an electron with its orbital motion. The orbital motion defines the electronic distribution around an atom, and this distribution can be probed with electric fields.

\begin{figure}[t]
%\begin{figure}[p]
\centering
\includegraphics[width=\columnwidth]{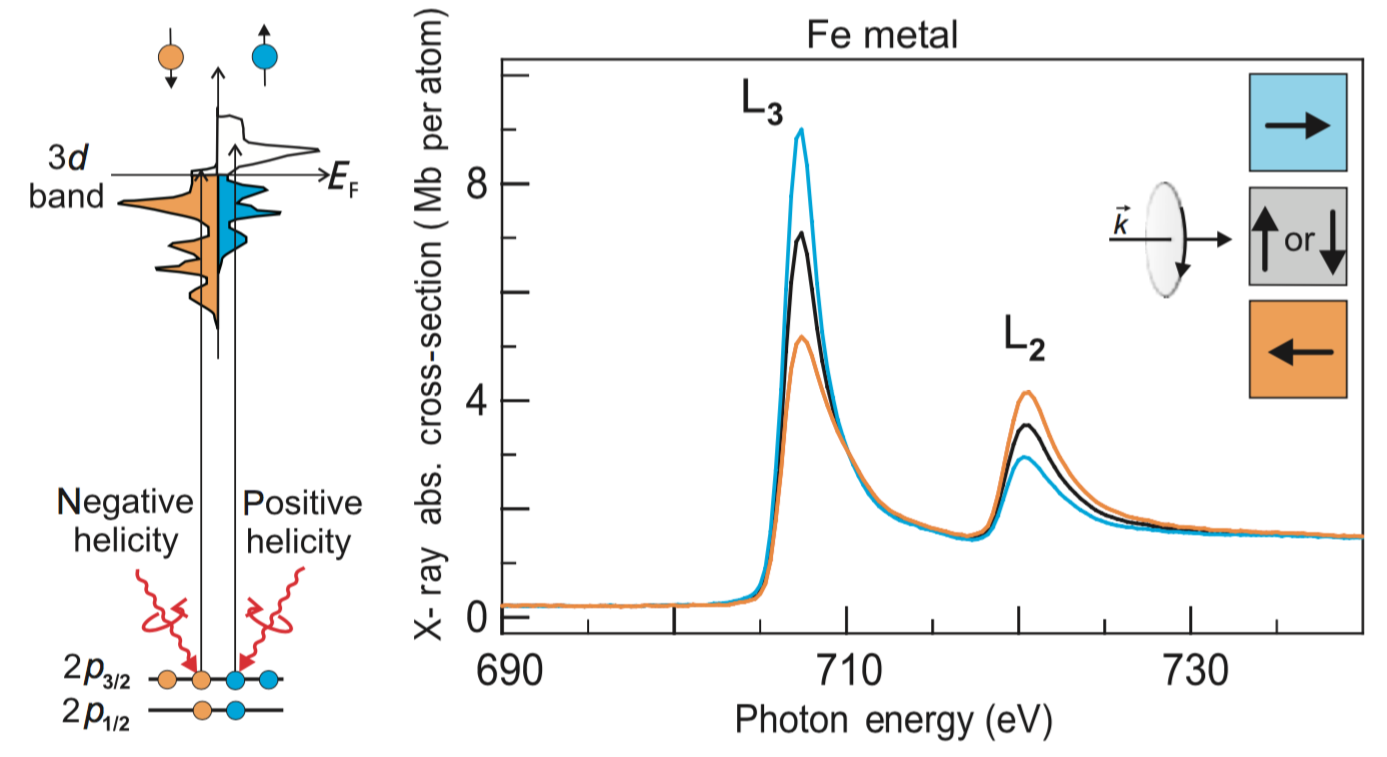}
\caption{Pictorial description of the mechanism of the X-ray magnetic circular dichroism (XMCD) effect. (Left) Schematic view of the two-step process that illustrates the different transition probability for two photons with opposite helicity tuned at the $L_2$ absorption edge. (Right) Experimental XMCD spectra of metallic Fe showing the two absorption peaks at the $L_3$ and $L_2$ edges for sample magnetised parallel (blue), antiparallel (orange) or orthogonal (black) compared with the X-ray propagation direction, as represented by the top-right inset. (From Ref.~\cite{stohr2007magnetism}, reprinted with permission.)}
\label{fig:xmcd}
\end{figure}
It is important to note that, in general, at X-ray frequencies, even the response of the electrons to electric fields is relatively modest, empirically described by the fact that the refractive index is 1 for all materials in the X-ray region. However, in the particular case of \emph{resonant} experiments, where one tunes the X-ray energy to be the same as the energy difference between two orbitals, the interaction is greatly amplified, and it becomes comparable to the strength of the interaction for visible radiation. For $3d$ elements, which include the room temperature ferromagnets Fe, Co and Ni, the important transition is the one between the $2p$ core levels and the $3d$ bands. If one measures the X-ray absorption around these transitions, two peaks corresponding to the $L_3$ ($2p_{3/2}$ to $3d$) and $L_2$ ($2p_{1/2}$ to $3d$) absorption edges can be observed, as shown in Fig.~\ref{fig:xmcd} for the case of Fe. It is around these two transitions that the effect of the spin-orbit coupling is strongest, and the XMCD can be observed.\footnote{Generally, the X-ray spectral range where XMCD is typically measured in $3d$ transition metals is called the \emph{soft} X-ray region. It is conventionally assumed to extend from the carbon $K$-edge ($E\approx284$ eV) to the copper $L_2$-edge ($E\approx952$ eV). The $L_3$ edges for the three magnetic elements, Fe, Co and Ni, are at $E_{\rm{Fe},L_3}\approx707$ eV, $E_{\rm{Co},L_3}\approx778$ eV and $E_{\rm{Ni},L_3}\approx852$ eV. Many soft X-ray beamlines at modern synchrotrons also provide X-rays over a slightly wider spectrum, and reach the $M$-edges ($3d$ to $4f$ transitions) of Gd ($E_{\rm{Gd},M_5}\approx1190$ eV) and Tb ($E_{\rm{Tb},M_5}\approx1241$ eV), rare earth elements that become strongly magnetic in proximity of $3d$ transition metals and that are interesting from a fundamental and applied perspective.}

In its simplest phenomenological explanation, XMCD is the contrast in X-ray absorption observed when the magnetisation of a sample is switched. To observe such contrast, besides being tuned to an absorption edge, the X-rays need to be circularly polarised, and there must be a component of the magnetisation vector that is collinear to the X-ray propagation direction. One way to memorise this is to think to the circularly polarised X-rays as photons with a spin: when the photon spin is parallel to the spins in the material, one observes enhanced absorption; when it is antiparallel, absorption is reduced. Photon spin or spins in the material can be equivalently switched to observe the same contrast. Accordingly, the baseline absorption is measured in samples with no-net magnetisation (or with magnetisation orthogonal to the X-ray propagation direction), or using with linearly polarised X-rays, as shown on the right side of Fig.~\ref{fig:xmcd}. Often, this contrast mechanism is all that one needs to understand the majority of imaging experiments.

An important point that needs to be made is that by using the XMCD contrast at the $L-$edges, one always probes the atomic, or the localised, magnetic moment. Therefore, strictly speaking, one cannot detect a spin \emph{current} with X-rays, but only an \emph{accumulation} of spins, or an induced magnetic moment in proximity of an interface. However, since the detection of a spin accumulation is associated with the presence of a spin current, the two concepts are sometimes exchanged.

Given that the XMCD probes the atomic moment, how does the contrast takes place at the microscopic level? This can be shown using the two-step model described in Ref.~\cite{stohr2007magnetism}. The first step is the absorption of the circularly polarised X-ray photon through the creation of a photoelectron. Energy is conserved by adding an electron to an empty valence state, although the circular polarisation of the photon also requires conservation of angular momentum. In general, this is satisfied by the photoelectron carrying the orbital angular momentum. In the presence of the spin-orbit coupling in the core levels, the photon angular momentum is also transferred to the spin of the photoelectron, which becomes spin-polarised. The quantization axis is given by the \emph{direction} of the photon spin, and the sign of the spin polarisation is given by the \emph{sign} ($\pm\hbar$) of the photon spin and by the sign of the spin orbit coupling, which is opposite for the $2p_{3/2}$ and $2p_{1/2}$ levels.

The second step in the model is the detection of the spin-polarised photoelectron in the $3d$ band. As we saw before, in a ferromagnet, the $3d$ band is divided into two bands for the majority and minority spins. Those are split by the exchange interaction, which creates an unequal population of spin-up and spin-down electrons and, conversely, of available states with opposite spins. Since the probability of a transition depends on the availability of empty states for a photoelectron, photoelectrons with opposite spin-polarisation are absorbed at different rates. In a typical imaging experiment where the energy and the helicity of the X-rays are fixed (i.e. the number and the spin-polarisation of the excited photoelectrons are fixed), the relative orientation of the spins in the sample, which defines the spin quantization axis and therefore the ``efficiency'' of the $3d$ bands as detectors, determines the overall absorbed X-ray intensity. This is the observed XMCD contrast.

\section{Scanning transmission X-ray microscopy at synchrotron light sources}
\begin{figure*}[t]
%\begin{figure}[p]
\centering
\includegraphics[width=\textwidth]{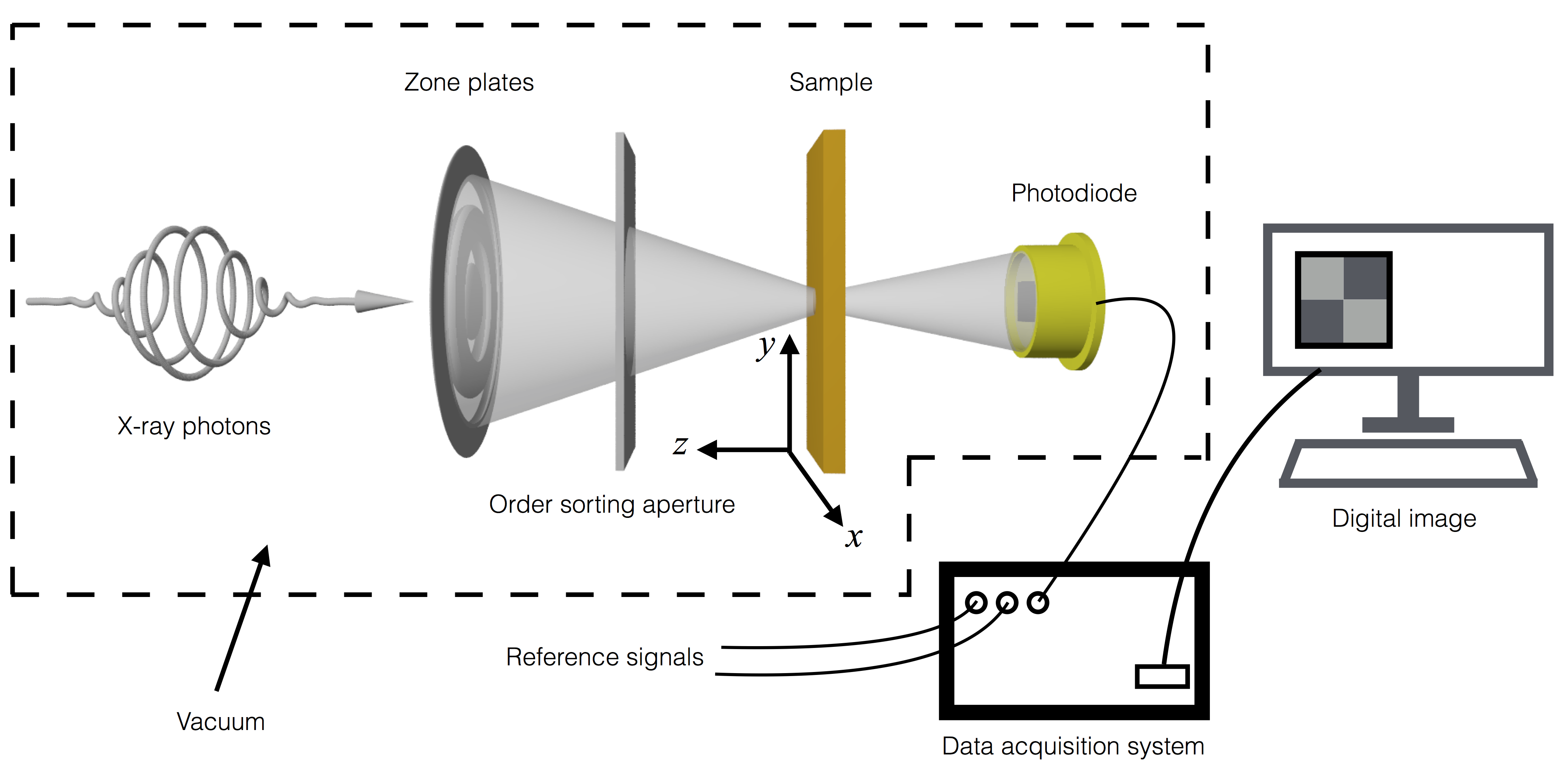}
\caption{Schematic of a STXM setup available at modern synchrotron light sources. The X-ray photons are produced by the deflection of $\sim50$ ps long electron bunches separated by a few nanoseconds that travel around in the storage ring. Typical operation frequencies of the storage rings are in the hundreds of MHz, and revolution frequencies are in the $\sim1$ MHz range. Timing signals are available at the beamlines requiring them.}
\label{fig:stxm}
\end{figure*}

A scanning transmission X-ray microscope \cite{rarback1988scanning,jacobsen1991diffraction,fischer2015x} is one of the conceptually simplest and most direct microscopy techniques. An X-ray beam is focused and sent through the sample, and the transmitted X-ray light is detected with a photodiode. An image is built by scanning the sample, which is mounted on movable stages, in the plan perpendicular to the X-ray propagation direction. In other words, the image is constructed pixel-by-pixel from the intensity recorded by the photodiode. A schematic of the setup is shown in Fig.~\ref{fig:stxm}.

Focusing the X-rays cannot be realised with conventional lenses, as the refractive index of all materials is very close to 1. Instead, Fresnel zone plates are used, which achieve focusing by constructive interference of the X-ray beam. Zone plates are made of concentric rings (the ``zones'') of opaque and transparent material, typically gold and silicon nitride, respectively. In order to achieve constructive interference, the radii $r_n$ of the edges (i.e. the boundaries between the opaque and transparent zones) need to decrease as the distance from the centre of the zone plates increases \cite{kirz1986x}:
\begin{equation}
r_n^2 = nf\lambda + n^2\lambda^2/4,
\end{equation}
where $n$ is the zone number (to be counted separately for opaque and transparent zones), $\lambda$ is the X-ray wavelength and $f$ is the first-order focus. In the long focal length limit, only the first term in the right-hand side of the equation is necessary. In this limit, the maximum possible resolution $\Delta l$ is given by:
\begin{equation}
\Delta l= 1.22\Delta r_N,
\end{equation}
where $\Delta r_N$ is the width of the outermost zone plate. This follows from the Rayleigh criterion $\Delta l = 0.61\lambda/NA$, where the numerical aperture is $NA=\lambda/2\Delta r_N$
\cite{attwood2007soft}. Standard commercial waveplates have a resolution of about 25 nm, and prototypes providing $\sim10$ nm resolution have been demonstrated \cite{chao2009demonstration}.

Typically, only $\sim10\%$ of the light is focused into the first-order focus by the zone plates, with the remaining X-rays being transmitted unfocused (zero-order radiation). To get rid of this part of X-rays, which would otherwise mask the signal from the focus, a precise pin-hole structure, called an order sorting aperture (OSA), is placed in front of the zone plates.

There are many advantages with this type of microscope. \emph{i)} No or minimal data processing is required to retrieve the data, which means that an image can often be observed on a computer screen as it is effectively being measured. \emph{ii)} It is a photon-in/photon-out technique, which allows for great versatility in terms of the sample environment. Electric and magnetic fields of, in principle, arbitrary magnitude can be applied to the sample, in contrast with techniques that are based on electron yield. Ultra-high vacuum is not required, although a good vacuum helps reduce the sample contamination, often in terms of carbon deposition. \emph{iii)} The use of a point detector also allows for fast gating of the signal, which is key to implementing time-resolved capabilities. Even at the high operation frequencies of synchrotrons, which produce $\sim50$ ps X-ray bursts at $\sim500$ MHz rate, detection of the individual bursts is possible using fast avalanche photodiodes (APDs) and advanced photon counting methods \cite{RSI.78.014702}.

There are two main issues that need to be addressed when using STXM: drift and sample fabrication. Drift is naturally present because the technique relies on the mechanical translation of stages over micrometre distances while requiring $\sim$$10$ nm accuracy. However, the reproducibility of such motion is still not good enough using purely mechanical means. Modern commercial STXM setups implement a laser interferometer mounted on the stages that allows for highly reproducible motions \cite{kilcoyne2003interferometer}, which solves this problem. However, sample fabrication is still only a partially solved issue. In order for the X-rays to be transmitted to the APD, samples need to be prepared on chips made of suspended SiN membranes of $\sim100$ nm thickness and $\sim100$ $\mu$m in lateral size. While those membranes are commercially available, only a subset of materials can be grown on them. In particular, it is still a challenge to prepare samples with crystalline order. Luckily, magnetic thin film samples, used in both applied and fundamental research, are often polycrystalline, which makes them compatible with the growth on SiN membranes.

Transient magnetic signals in nanometre-thick magnetic films may result in a small variation of the total X-ray flux, down to a $10^{-4}-10^{-5}$ relative signal change. This poses important challenges in three domains of the measurements. First, the reliable detection of the X-rays transmitted through the sample. When measuring the photons coming from individual synchrotron bunches, one often ends up in the single-photon regime, which requires advanced detection methods \cite{RSI.78.014702,Kaznatcheev:AipConferenceProceedings:2011}. Second, the stable synchronisation between the signals exciting the sample (typically electrical signals) and the detector \cite{strachan2007synchronized}. Such synchronisation allows for stroboscopic detection, which is necessary to avoid washing out the signal and is key to reducing the averaging time per image. Third, the correction for all drift that naturally occurs when measurement times are of the order of several hours, as is often the case when one has to achieve an acceptable signal-to-noise ratio and the signals are small. Since such drift cannot be completely removed in large facilities affected by daily thermal excursions and tidal forces, one possible solution is to have a normalisation signal that follows the very same drift. For example, one way to achieve this in pump/probe experiments is to excite the sample only during even revolutions of the electron orbit, and to use the odd revolutions to record the normalising images. This has also the advantage to exactly cancel out the nonlinearities in the detector, which is physically the same for both the reference and the actual signal. This clearly requires the ability to gate both the detection and the excitation signals at the revolution frequency, typically of the order of 1 MHz, but this can often be achieved by commercial electronics. The details on how to implement this to be able to record STXM images up to 10 GHz rates can be found in Ref.~\cite{bonetti2015microwave}.

\section{Imaging spin currents}
In this section, we summarise the recent experimental efforts that have been successful in directly mapping the presence of spin currents in magnetic nanostructures. First, we describe how STXM has been used to record images of spin transfer torque switching events in nanomagnets with different anisotropies. Then, we discuss how the same technique has been successfully implemented to directly image spin accumulation in non-magnetic Cu, effectively detecting the flow of a spin current.

\subsection{Spin torque induced switching}
\begin{figure}[t]
%\begin{figure}[p]
\centering
\includegraphics[width=\columnwidth]{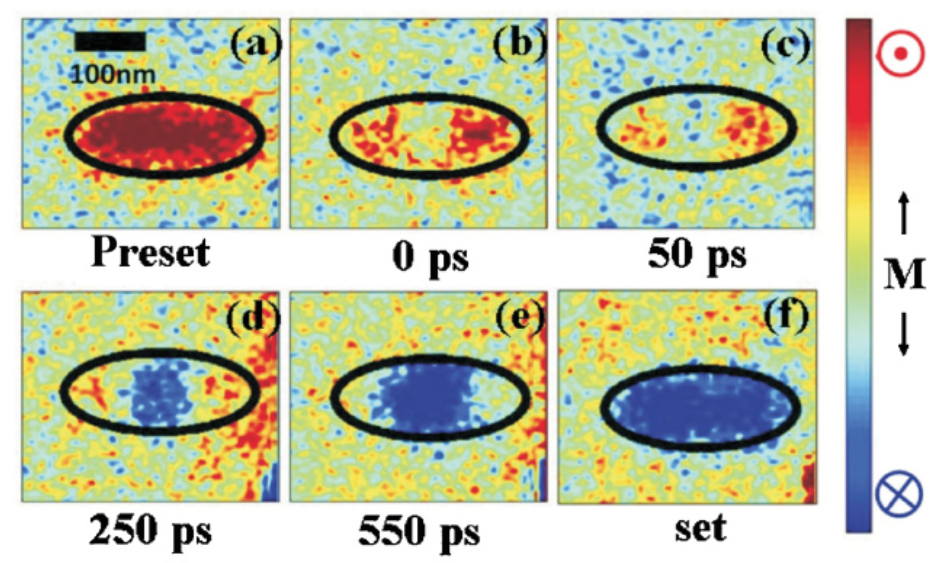}
\caption{Temporal snapshots of spin torque induced switching in a magnetic spin valve with perpendicular magnetic anisotropy. The colour scale indicates the component of the magnetisation perpendicular to the film and the image plane. The six images are recorded at different time intervals from the arrival of a current pulse, indicated below each snapshot. (From Ref. \cite{bernstein2011nonuniform}, reprinted with permission.)}
\label{fig:STT_switching}
\end{figure}
The observation of current-induced magnetisation switching in a patterned spin valve \cite{myers1999current} was the first experimental verification of the correctness of Slonczewski prediction: that spin transfer torque can be large enough to cause a reversal of the magnetisation towards a new equilibrium state. However, such observation was indirect: the reversal was detected as a change in the overall resistance of the device through the giant magnetoresistance effect.

Therefore, it was highly desirable to directly observe the effect of the spin current on the nanomagnet, and to directly image the switching of the magnetisation as it happens. A breakthrough happened in 2006 when Acremann and co-authors were able to map the switching of the ``free'' layer of an in-plane magnetised spin valve \cite{acremann2006time} using the electrical pump and X-ray probe detection technique described above. They made a surprising observation: the Oersted field, which was typically assumed to be negligible compared with the spin transfer torque in structures at the $\sim$$100$ nm scales, turned out to be of fundamental importance. In fact, the Oersted field creates an asymmetric potential well (when an in-plane field is applied to the sample), around which switching occurs. A subsequent X-ray study in combination with micromagnetic simulations could reveal that the switching occurs through the motion of a magnetic vortex \cite{strachan2008direct}. These observations were key to understanding the mechanism of magnetisation switching at the nanoscale, which can be better controlled by playing with the magnetic anisotropies in the system. 

A similarly surprising observation was made in a subsequent work \cite{bernstein2011nonuniform} using a sample with perpendicular magnetic anisotropy (PMA), i.e. where the easy axis of magnetisation is in the out-of-plane direction with respect to the film plane. In this case, the Oersted field had cylindrical symmetry, as the applied field had no in-plane component that could break the in-plane symmetry. However, even in this case, the spin reversal was not uniform, as shown in Fig.~\ref{fig:STT_switching}. One can clearly see that in the elliptical nanostructure, the reversal is initiated in its central part, and it propagates outwards at a later stage.

The shared message from these two works, obtained thanks to the unique capabilities of x-ray microscopy, is that in magnetisation reversal, spatial inhomogeneities are fundamental even at the nanometre scale, and that the often used approximation of single-domain, breaks down dramatically.

\subsection{Direct imaging of spin currents in non-magnetic materials}

\begin{figure}[t]
%\begin{figure}[p]
\centering
\includegraphics[width=\columnwidth]{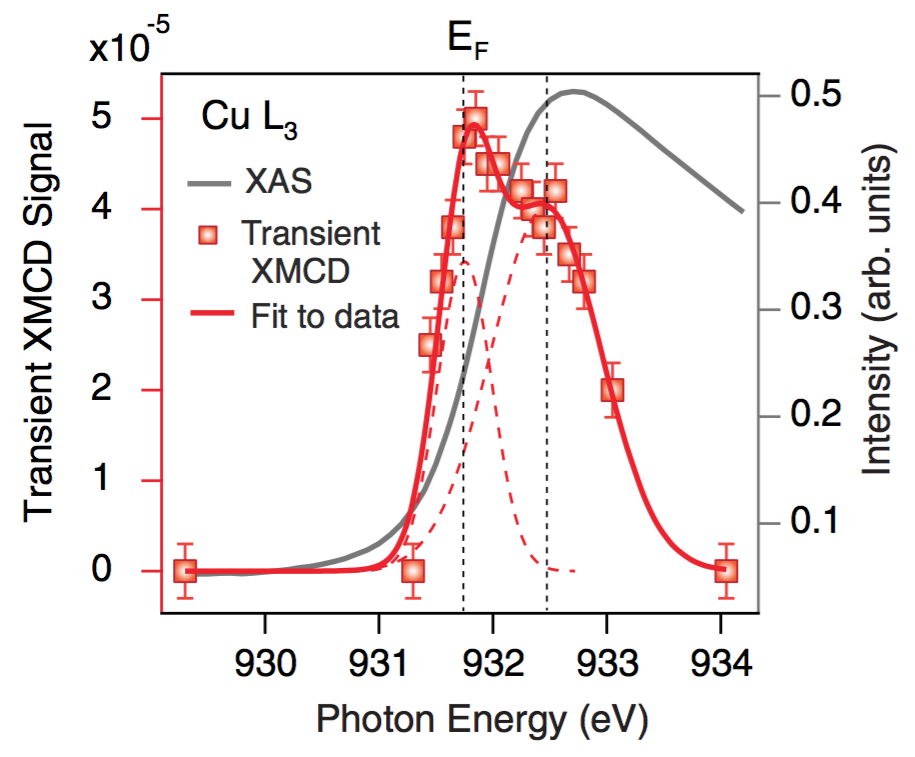}
\caption{Absorption spectrum of Cu in proximity of the $L_3$ edge (gray line) and the Cu XMCD signal (red squares) across a Co/Cu interface traversed by an electrical current. The signal is present only in the transient time when the current is switched on. The transient spectrum (red line) can be fitted by two Gaussians (dashed red). (From Ref. \cite{kukreja2015x}, reprinted with permission.)}
\label{fig:spin_injection}
\end{figure}

Having succeeded in imaging spin reversal in a ferromagnet, the next challenge was to directly observe the spin current that causes the reversal. As we discussed earlier, we cannot directly observe the flow of a spin, but we can look for the appearance of spin accumulation across a ferromagnet/non-magnet interface. As we explained in Section 2, the build up of spin accumulation is a fingerprint of a spin current flowing through an interface.

Observing the occurrence of spin accumulation is, however, extremely challenging, as the associated magnetic moment is extremely tiny. For a nanopillar, where the current flows across the layers, the magnetic moment can be estimated as \cite{stohr2007magnetism}:
\begin{equation}
m=-D(E_F)\left[\mu^\uparrow(0)-\mu^\downarrow(0)\right]\frac{1}{d}\mu_B\int_0^d e^{-x/\Lambda}dx,
\end{equation}
where $D(E_F)$ is the density of states at the Fermi energy, $\mu^{\uparrow(\downarrow)}(0)$ is the chemical potential at the interface ($d=0$) for the majority (minority) spins, $d$ is the thickness of the layer and $\Lambda$ is the spin diffusion length. Calculations using realistic parameters for a Co/Cu interface predict $m_{\rm{Cu}}\approx9\times10^{-5}$ $\mu_B$/atom.

The first experimental proof that a spin accumulation signal was tiny came from the experiment by Mosendz \emph{et al.} \cite{mosendz2009imaging}. They used a full-field microscope to image the spins injected from a Co film into a Cu wire in a lateral spin valve. The authors could not observe a signal, but were able to establish an upper limit for the signal to be detected. According to their calculations, spin accumulation in Cu should result in a variation of the XMCD signal of the order of $10^{-4}$.

The experimental confirmation of both the theoretical and experimental predictions happened only very recently, when spin accumulation at a Co/Cu interface could be detected spectroscopically using STXM \cite{kukreja2015x}. These measurements are summarised in Fig.~\ref{fig:spin_injection}. The transient XMCD signal caused by the flow of current resulted in a relative variation of $5\times10^{-5}$ over the background X-ray transmission, for the photon energies around the inflection point of the absorption peak. These photon energies are the ones that induce a transition from the $2p$ core level to the Fermi energy $E_F$, with the broadening of the peak due to the finite energy resolution of the beamline. The relative variation of the XMCD signal could then be converted to a transient magnetisation of $\sim3\times10^{-5}$ $\mu_B$ per Cu atom at $E_F$, within a factor of 3 of the theoretical value \footnote{It can be shown that the detected signal corresponds to the XMCD signal that would be generated by $\sim50$ Fe atoms.}.

While the Fermi level data were consistent with well-established theories and transport measurements, the spectroscopic data away from the Fermi level showed an unexpected feature. Roughly 1 eV above the Fermi level, a second peak could be clearly distinguished. Since the peak was at a relatively large distance from the Fermi energy, it could not be explained in terms of scattering of conduction electrons in the bulk, with energies that are within a few meV from the Fermi level. Instead, the peak was found in correspondence of the Fermi energy of the interface, where the Cu and Co energy bands hybridize and where the magnetic moment is reduced. The interpretation given to this observation was that part of the spins polarised in the Co layer, while flowing through the interface, got ``stuck'' as they transferred angular momentum (via spin torque) to the interface, which had lost part of its magnetisation due to hybridization. The quantitative considerations made possible by XMCD showed that a couple of atomic layers could accumulate as many spins as $\sim$$30$ nm of bulk Cu. This means that roughly half of the spins were blocked at the interface. This information, which is extremely challenging to obtain with conventional transport measurements, may be key to better understanding spin-dependent scattering in real interfaces and designing efficient spin injectors.

Very recently, another experimental study reported on the observation of a pure AC spin current in a metallic permalloy/Cu/Cu$_{75}$Mn$_{25}$/Cu/Co multilayer \cite{PhysRevLett.117.076602}. In this work, the excitation of the ferromagnetic resonance in the permalloy thin film was used to generate a spin current (via the spin-pumping mechanism \cite{tserkovnyak2002enhanced}) in the Cu$_{75}$Mn$_{25}$ layer. The spin current was detected as a temporal variation of the XMCD signal at the Mn $L_3$ edge, and measured at the same frequency of the ferromagnetic resonance in the permalloy layer. By replacing the first Cu layer with an insulating MgO layer, the spin current was effectively blocked and no variation in the XMCD signal at the Mn edge was observed. The authors also showed that the generated spin current could be used to transfer a spin torque on the Co layer, demonstrating the flow of net angular momentum across the multilayer.

\section{Imaging magnetisation dynamics}

In this section, we discuss recent experimental efforts in imaging magnetisation dynamics at the nanoscale. Interest in using X-rays to look at magnetisation dynamics was first stimulated by investigations on magnetic multilayers where the dynamics between the layers is coupled \cite{bonfim2001element,vogel2003time,bailey2004precessional,arena2006weakly,guan2006comparison, arena2007combined, Guan:JournalOfMagnetismAndMagneticMaterials:2007, martin2008layer, martin2009layer, Arena:ReviewOfScientificInstruments:2009,Boero:RevSciInstrum:2009,Goulon:IntJMolSci:2011, Cheng:JournalOfAppliedPhysics:2012, Rogalev:JournalOfInfraredMillimeterAndTerahertzWaves:2012, Warnicke:JournalOfAppliedPhysics:2013, bailey2013detection, stenning2015magnetization}. One of the unique characteristics of X-rays is that they allow for element-specific detection of the magnetic signal. This can be simply achieved by tuning their energy, which can therefore allow individual layers to be probed. These efforts led to refined schemes for detecting time-resolved XMCD signals, which were then implemented in STXM to \emph{image} the dynamics.

In recent years, interest in the control of spin waves in magnetic nanostructures has increased dramatically, becoming the focus of the research field of \emph{magnonics} \cite{neusser2009magnonics,kruglyak2010magnonics}. The ultimate goal of magnonics is to realise circuits where spin waves (i.e. \emph{magnons}) are used to generate, control and detect information \cite{bonetti2013nano} in magnetic nanostructures. In these structures, the confinement introduces additional features to the spin wave spectrum, such as the appearance of quantized and edge modes \cite{demokritov2009spin, cheng2017phase}. All these aspects can be understood only partially using conventional FMR methods, while imaging techniques are required to retrieve the full picture of the physics at play. The work horse in these cases is the technique of microfocused Brillouin light scattering \cite{demidov2004radiation,sebastian2015micro}, which is able to provide spatial maps of magnetisation dynamics without the need of synchrotron radiation. However, the relatively limited spatial resolution of the technique (typically $\sim$$250$ nm), as well as the impossibility of layer-resolved measurements, are sometimes critical obstacles in nanostructures of fundamental and applied interest. Time-resolved X-ray microscopy does not have such limitations and can be used to image FMR and spin wave modes with extremely high-resolution.

We will discuss imaging experiments of magnetisation dynamics in patterned magnetic structures excited by AC magnetic fields, in particular vortex dynamics. We will then describe recent studies where the spin waves induced by spin transfer torque were directly probed with time-resolved microscopy.

\subsection{Magnetisation dynamics in confined structures: Vortices and skyrmions}
\begin{figure}[t]
%\begin{figure}[p]
\centering
\includegraphics[width=\columnwidth]{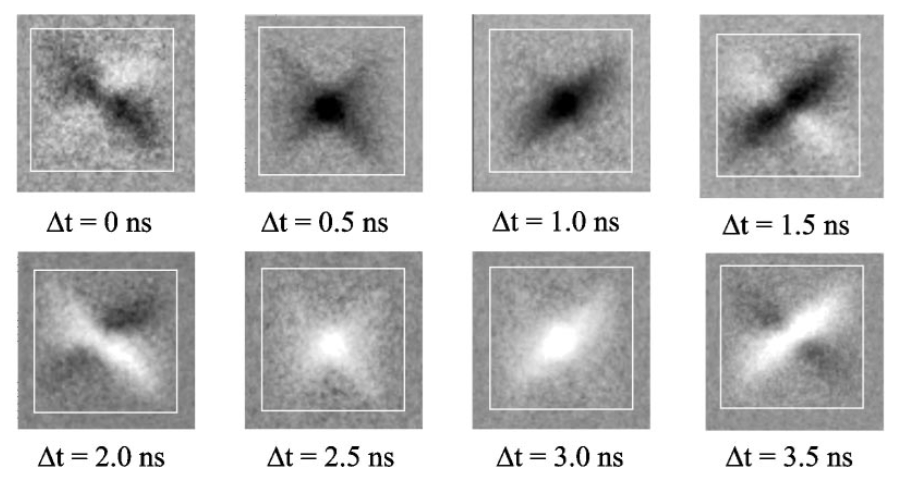}
\caption{Spatially and temporally resolved images of magnetisation dynamics in $1\times1$ $\mu$m$^2$ patterned structures recorded by time-resolved XMCD. The eight different panels represent eight different phases (45 degrees apart) of one precession period of 4 ns ($1/250$ MHz) of the dynamics. The white to black contrast is a measure of the magnetisation in the plane of the sample along the horizontal direction. The images can be analysed to reveal the presence of a magnetic vortex. (From Ref.~\cite{puzic2005spatially}, reprinted with permission.)}
\label{fig:xfmr}
\end{figure}

\begin{figure*}[t]
%\begin{figure}[p]
\centering
\includegraphics[width=\textwidth]{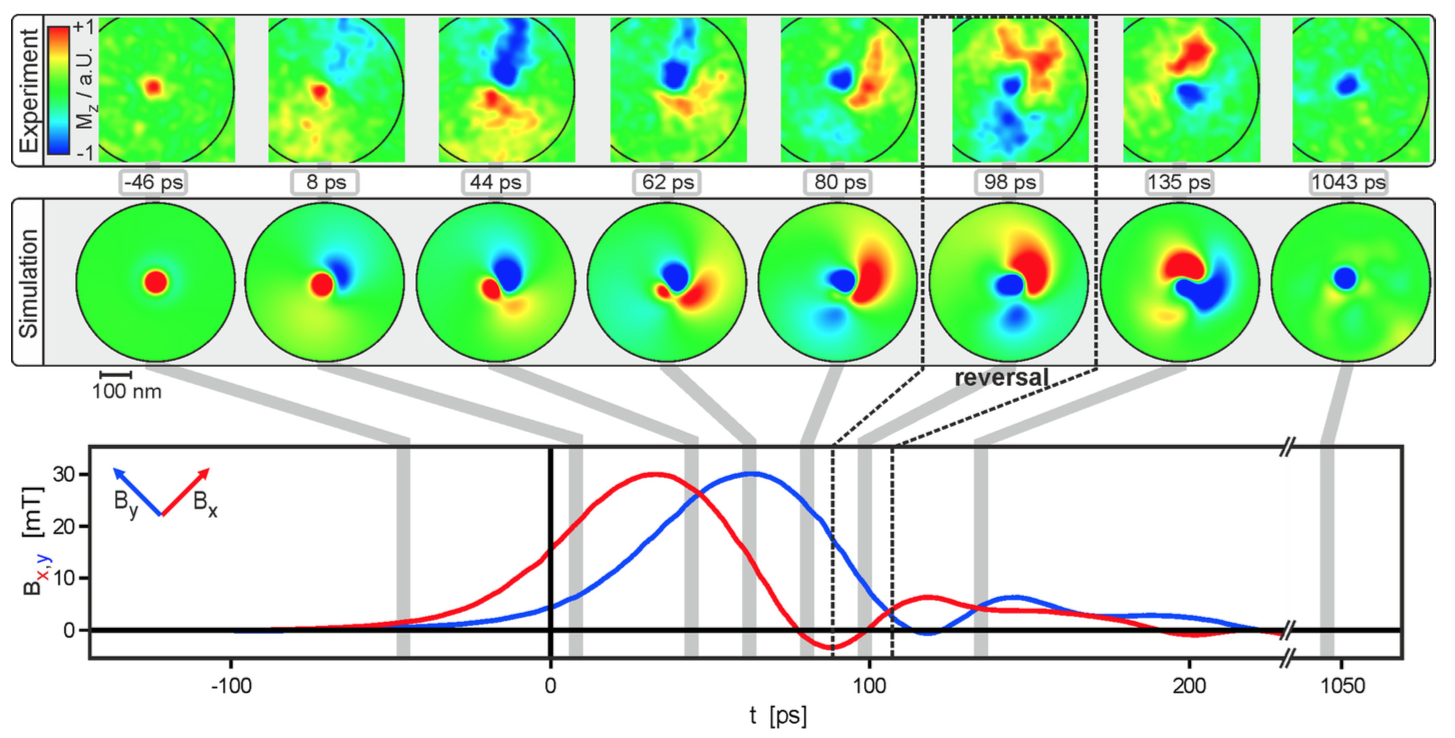}
\caption{(Top row) Time-resolved XMCD images of vortex dynamics triggered by pulsed magnetic fields in a $\sim50$-nm thick, micrometre-sized permalloy (Ni$_{80}$Fe$_{20}$) disks. The blue to red contrast measures the amplitude of the magnetisation perpendicular to the thickness of the permalloy and to the image plane. (Middle row) Dynamics of the out-of-plane component of the magnetisation computed by micromagnetic simulations using the same parameters of the experiment. (Bottom row) Temporal profile of the magnetic field $B$ ($x$ and $y$ components in red and blue, respectively) that triggers the dynamics. (From Ref.~\cite{noske2014unidirectional}, Creative Commons.)}
\label{fig:vortex_sw}
\end{figure*}

The dynamics of spin structures in confined geometries is distinct from continuous films, as the geometrical confinement leads to novel spin structures that because of their topology exhibit particular dynamics. In general, shape anisotropy leads to vortex states in discs \cite{shinjo2000magnetic}, to geometrically-defined domain walls in wires \cite{Boulle:MaterialsScienceAndEngineeringRReports:2011} and to single skyrmions and geometrically confined skyrmion lattices in perpendicular anisotropy materials \cite{woo2016observation}.

A particularly interesting spin structure is the vortex. Pioneering experiments in imaging vortex dynamics have been conducted by Stoll and co-authors \cite{stoll2004high}, Choe \emph{et al.} \cite{Choe:Science:2004} and Puzic \emph{et al.} \cite{puzic2005spatially}
using full field X-ray microscopy, the latter experiment being summarised in Fig.~\ref{fig:xfmr}. The authors applied a 250 MHz rf field to a $1\times1$ $\mu$m$^2$ Ni$_{80}$Fe$_{20}$ (permalloy) structure that was magnetised in-plane. The sample was tilted with respect to the plane perpendicular to the X-ray photons so that there was a finite component of the magnetisation along the X-ray propagation direction, which was necessary to detect an XMCD contrast. Such contrast is presented as a gray scale variation for eight different phases of the precession. The resolution of the image was not explicitly measured in that work, but it is clearly well below 100 nm, and likely limited by the sharpness of the magnetic features. By analysing these spatially and temporally resolved maps, the authors could explain the features of the images as a consequence of the gyration of a magnetic vortex core.

After those pioneering imaging experiments, other investigations have been performed using a similar experimental approach. An important success in time-resolved X-ray imagining with STXM, which demonstrated the power of the technique, was the discovery that a vortex core reverses through the excitation of the vortex gyrotropic mode and through the creation of a vortex-antivortex pair \cite{van2006magnetic}. The intrinsic elemental selectivity of X-rays was then used to study the interaction between vortices in coupled magnetic multilayers \cite{chou2006vortex}. STXM also allowed researchers to observe the suppression and nucleation of a vortex in proximity of artificially induced defects \cite{kuepper2007vortex}. Subsequent time-resolved X-ray imaging experiments have been used to demonstrate that the polarity of a vortex core can selectively be switched with rotating magnetic fields of opposite helicity \cite{curcic2008polarization}, and to image the oscillation of domain walls \cite{bocklage2008time}, as well as stochastic domain-wall depinning in nanowires \cite{im2009direct}. Similar studies were extended towards the understanding of the dynamics of coupled structures, and the role of the shape of the magnetic structures in the coupling \cite{jung2011tunable,vogel2011coupled}. In the limit of large-amplitude driving rf magnetic fields, it was also shown that it is possible to observe non-equilibrium dynamic configurations of Landau states in a square permalloy disk that contains a vortex \cite{stevenson2013dynamic}. Those excitations were identified as the spatial-domain equivalent of the frequency-domain Suhl instabilities.

Recent studies have been able to map the GHz magnetic susceptibility in micrometre-sized structures \cite{APL.101.182407}, identify a correlation between the velocities of domain walls and the transformations of the spin structure \cite{bisig2013correlation} and observe the synchronous motion of multiple domain walls in nanowires \cite{kim2014synchronous} and the stochastic formation of vortices \cite{im2014stochastic}. It has also been shown that the polarisation of a vortex core in a permalloy film can be switched when crossing a domain wall of an adjacent magnetic film with out-of-plane anisotropy in a maze domain state \cite{wohlhuter2015nanoscale}. The possibility of imaging the dynamics of the vortex domain wall nucleation \cite{richter2016localized} and the wall propagation dynamics in asymmetric rings \cite{richter2016local} also show the importance of the local geometry that governs the spin structure dynamics and can be used to tailor it.

Further improvement to the detection sensitivity and the compatibility of STXM with high-frequency electronics led to the achievement of spectacular time-resolved imaging capabilities, as illustrated by the work of Kammerer \emph{et al.}; in this study, they could observe the details of the mechanism of switching of a magnetic vortex caused by the formation of a vortex--antivortex pair and subsequent vortex--antivortex annihilation \cite{kammerer2011magnetic}. In follow-up works, the same experimental team was able to observe further  details of the interactions between spin waves and vortices \cite{kammerer2012fast,sproll2014low,bauer2014vortex,noske2014unidirectional}. An example of the images that can be recorded is shown in Fig.~\ref{fig:vortex_sw}. The top row of the figure shows the experimental images, while the middle row the micromagnetic simulations. The agreement between the two series of images is outstanding. A recent and detailed review on the topic of vortex dynamics imaged by time-resolved XMCD can be found in Ref.~\cite{stoll2015imaging}.

In more recent works, it was demonstrated that the vortex core dynamics in antiferromagnetically coupled layers can be used to create spin waves with a wavelength that can be tuned linearly with the frequency of the driving signal \cite{wintz2016magnetic}, and that collective modes can be observed in arrays of permalloy disks containing magnetic vortices forming a so-called ``vortex crystal'' \cite{hanze2016collective}.

Vortex dynamics can be excited not only by external rf magnetic fields, but also by spin transfer torque that can drive large-amplitude, nonlinear vortex dynamics in confined structures. Even in this case, time-resolved XMCD microscopy can be used to image such dynamics directly. The first experimental observation of vortex gyration that could be partially explained by spin transfer torque was reported for a permalloy square with lateral contacts feeding an alternating current to the magnetic structure at a frequency of about 60 MHz \cite{bolte2008time}. In 2011, Yu \emph{et al.} managed to capture, for the first time, the dynamics of a magnetic vortex in a nanopillar driven exclusively by the spin torque provided by the direct current through the nanopillar \cite{PhysRevLett.106.167202}. (An alternate current, two orders of magnitude smaller than the direct current, was used only to phase lock the excitation with the synchrotron frequency.) By achieving a resolution of 30 nm, the authors could record a movie of the vortex dynamics and realise that such dynamics was substantially more complicated than usually assumed. In order to perform such an experiment, a driving signal at a frequency as high as 1 GHz was required, and the authors further developed the dedicated pump-probe scheme connected to the STXM described above \cite{RSI.78.014702,strachan2007synchronized}.

Finally, beyond the formation of vortices that primarily occurs in soft magnetic materials as a result of dipolar interactions, qualitatively different spin structures can form in high anisotropy materials. These structures, such as skyrmions that can be stabilised by additional chiral exchange interactions, can also be imaged using X-ray microscopy \cite{woo2016observation}.

\subsection{Spin waves in spin torque oscillators}

\begin{figure*}[t]
%\begin{figure}[p]
\centering
\includegraphics[width=0.9\textwidth]{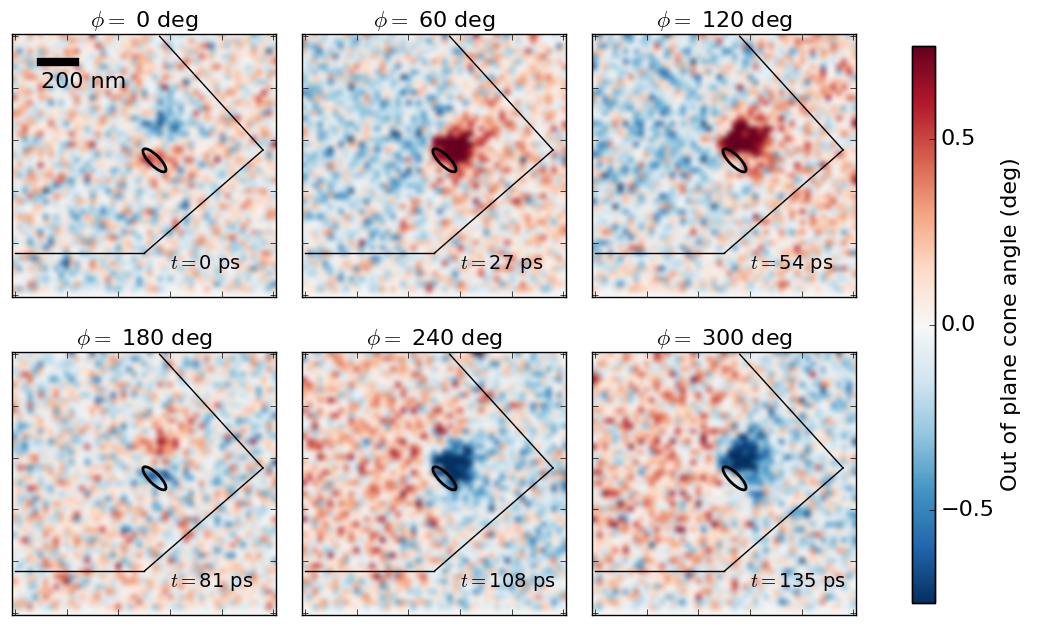}
\caption{Time-resolved XMCD images of spin wave dynamics excited in an extended permalloy layer by a spin-polarised current injected through a nanocontact (black ellipse). The six images are snapshots at six different phases of the oscillation with a precession period of $\sim160$ ps. The film is in-plane magnetised when no current is applied, and the presence of a contrast (blue to red) in the images is proportional to the out-of-plane component of the magnetisation that is created by the spin transfer torque. The centre of mass of the spin wave is offset by the centre of the nanocontact because of the magnetic potential minimum created by the superposition of the Oersted field created by the current flowing in the plane perpendicular to the image and an applied static magnetic field along the horizontal direction. The thin black lines mark the position of the electrodes, whose topographical features are removed to isolate the XMCD contrast. (From Ref.~\cite{Bonetti:NatureCommunications:2015}, Creative Commons.)}
\label{fig:sw}
\end{figure*}

Since the early works on spin torque induced dynamics, researchers have wondered about the spatial characteristics of the spin waves emitted by a nanocontact into extended magnetic films. Slonczewski originally predicted that an out-of-plane magnetised thin film would emit spin waves with a propagating character, with a wavelength $\lambda=2\pi(R_c/1.2)$, where $R_c$ is the radius of the nanocontact \cite{slonczewski1999excitation}. A few years later, Slavin and Tyberkevich \cite{slavin2005spin} predicted that for an in-plane magnetised nanocontact, a localised soliton, so-called spin-wave bullet, would more favourably form underneath the nanocontact. Micromagnetic simulations were able to prove the existence of both modes, but experimental verifications were still lacking \cite{consolo2007excitation}. A few years later, a combined experimental/micromagnetic study \cite{bonetti2010experimental} demonstrated that an out-of-plane magnetised contact generates propagating waves until the equilibrium magnetisation falls below a critical angle $\theta_c$. Below such angle, both spin wave modes exist, but they are intermittently excited in time. Subsequent experiments confirmed this picture \cite{consolo2013non,dumas2013spin}.

Despite all these efforts, the experiments relied on indirect information retrieved by analysing the spin-wave emission spectra. Two breakthrough experiments were performed in which microfocused Brillouin Light Scattering ($\mu-$BLS) \cite{demidov2004radiation,sebastian2015micro} was used to directly look at the spatial characteristics of the spin wave. At first, Demidov \emph{et al.} \cite{demidov2010direct} recorded the first images of the spin waves emitted by an in-plane magnetised nanocontact, while Madami \emph{et al.} \cite{madami2011direct} were able to capture the BLS signal from an out-of-plane magnetised contact. However, the resolution of the BLS, diffraction-limited at approximately 250 nm, did not allow for a detailed imaging of the spin wave. Also, the relatively long integration times ($\sim$$1$ second) needed to detect the spin wave signal, intrinsic of this type of detection, did not allow for the observation of the time-structure of the spin wave.

Only very recently has it been possible to perform time-resolved XMCD experiments with the combined temporal and spatial resolution needed to record a ``movie'' of the spin wave dynamics at the nanoscale in an in-plane magnetised nanocontact \cite{Bonetti:NatureCommunications:2015}. One of the technical steps that had to be implemented was to extend the capabilities of the time-resolved XMCD up to the $\sim10$ GHz rate \cite{bonetti2015microwave}.

The results of this experiment are summarised in Fig.~\ref{fig:sw} where six snapshots of the magnetisation dynamics are plotted \cite{Bonetti:NatureCommunications:2015}. The colour scale represents the amplitude of the magnetisation component parallel to the X-ray propagation direction (orthogonal to the thin film plane in this case), which is translated in terms of the out-of-plane cone angle of the magnetisation vector. The sensitivity of the technique allows the detection of sub-degree angles, with a noise on the order of $\sim0.1$ degrees. In this case, the interval between two consecutive snapshots is 27 ps, which is shorter than the $\sim50$ ps FWHM of the X-ray pulses. This means that the measured signal, which is the convolution of the real signal with a Gaussian function (representing the X-ray pulse), is smaller than the actual signal. The signal-to-noise ratio is, however, large enough to allow for a successful detection of the magnetisation precession.

The most surprising result of this study, enabled by the combined temporal and spatial resolution, is that the spin-wave shows a $p$-like character during the precession. This is clearly observed in the plots at $\phi=0$ $\phi=180$ degrees in Fig.~\ref{fig:sw}, where the contrast shows a node (i.e. a white region) between two regions where the magnetisation points in opposite directions (red and blue). This fact was not predicted by previous theories or simulations, whereas a cylindrically symmetric mode, with a $s$-like character, was expected. However, accurate micromagnetic simulations performed with the same parameters as the experiment could reproduce the measured data very well. This study also demonstrated the clear advantage in being able to compare experiments and simulations not only in terms of frequencies associated with the precession of the spin-wave, but also with respect to its spatial profile.

We conclude this section by mentioning that another type of spin wave excitation has been predicted in thin films with perpendicular magnetic anisotropy: droplet solitons \cite{hoefer2010theory}. The theory predicts that a droplet soliton is a full reversal of the magnetisation below the nanocontact where the current is injected. Such droplets have been observed experimentally \cite{mohseni2013spin, macia2014stable}, but there are still questions on the details of these excitations that indirect transport measurements cannot unambiguously answer. Indeed, work that looked at such dynamics directly using magnetic X-ray microscopy showed that the amplitude of the excitation, directly accessible through normalisation to the static XMCD contrast, appears to be much smaller (approx. 25 deg) than the predicted full reversal \cite{backes2015direct}. These measurements showed that new, more accurate models are needed to describe spin dynamics in these cases, where the role of thermal fluctuations, disorder in the material and the possible presence of overlying lower-frequency excitations need to be included.

\section{Looking forward: Novel spin physics and new tools}
\subsection{Novel spin physics}

The spin Hall effect was predicted half a century ago, and it was observed soon after in semiconductors \cite{chazalviel1972experimental, chazalviel1975spin}. However, in metals, it has only recently been observed in experiments\cite{valenzuela2006direct}. It consists of the appearance of a spin current that is \emph{transverse} to the charge current, and it is particularly large in heavy metals with large spin-orbit coupling. In thin films, a charge current flowing in the plane of the material induces a spin current towards the interface of the film, which can in turn be used to manipulate the magnetisation of an adiacent ferromagnet through an interfacial torque. Spin Hall induced torques have been demonstrated with Pt \cite{liu2012current}, and giant effects observed with Ta \cite{liu2012spin} and W \cite{pai2012spin}. A comprehensive review of the spin Hall effect can be found in Ref.~\cite{sinova2015spin}. An effect related to the spin Hall effect is the flow of a fully polarised spin current around the edges of topological insulators \cite{zhang2009topological}. In this novel class of materials, a conducting surface state is formed at the boundary between the topologically insulating bulk state and the trivially insulating state of the vacuum (or of another non-topological material). The surface state is very robust because it is protected by the topology of the bands, and many remarkable properties arise from this. One property is the spin-momentum locking of the carriers in the surface state, which is the reason for the full spin polarisation. Recently, it was demonstrated that the spin current at the interface of a topological insulator can be used to exert a large torque on the magnetisation of an adjacent ferromagnet \cite{mellnik2014spin}, and that a magnetically doped topological insulator can be switched with electric fields \cite{fan2014magnetization, fan2016electric}.

Another possible way to manipulate spin currents without the need for magnetic fields is to use magneto-electric multiferroics \cite{spaldin2005renaissance}. In these materials, the ferroelectric and the ferromagnetic orders are coupled to each other, and the manipulation of the ferroelectric polarisation through electric fields affects the magnetisation, and vice versa. In noncollinear magnets, the formation of electric polarisation can be explained in terms of a spin current \cite{katsura2005spin}. Recently, an ultrafast manipulation of the magnetic order through electric fields has been demonstrated for the noncollinear magnetic structure \cite{kubacka2014large}.

Finally, it has been predicted \cite{battiato2010superdiffusive} that ultrafast spin currents can be created using femtosecond laser pulses that induce demagnetisation or magnetisation switching in thin film ferri- and ferromagnets \cite{beaurepaire1996ultrafast,stanciu2007all}. An observation consistent with the presence of such ultrafast spin currents has recently been made by measuring the resonant magnetic X-ray scattering of fs X-ray pulses \cite{graves2013nanoscale}, and other experiments seem to confirm this prediction \cite{rudolf2012ultrafast, choi2014spin}. However, a direct imaging of the effect is still lacking, and competing mechanisms have also been proposed \cite{malinowski2008control,tows2015many}. The development of imaging techniques in the ultrafast regime is expected to open up for time-resolved XMCD investigations with combined nm and fs resolution using lensless holographic techniques \cite{eisebitt2004lensless,wang2012femtosecond,von2014imaging, buttner2015dynamics}, which will likely contribute towards finding a solution to this outstanding problem.

\subsection{New X-ray lightsources: Towards ultrasensitive and ultrafast imaging}
On the instrumentation side, the future for X-ray science is extremely bright. Two complementary directions are being taken towards developing increasingly better X-ray sources: diffraction limited storage rings (DLSRs) at synchrotron facilities and free electron lasers (FELs).

At the synchrotron facilities, accelerator physicists have been able to constantly reduce the emittance and to increase the brightness of the radiation over the past two decades. However, a quantum leap happened a decade ago, when a novel design of the storage ring optics, the multibend achromat, was introduced \cite{eriksson2008some}. The new design allows for both an order of magnitude reduction in the emittance and for a much more compact synchrotron design, which also greatly reduces the construction costs. The first DLSR synchrotron lightsource based on this design, the MAX IV Laboratory, has just started operation in Lund, Sweden \cite{eriksson2014diffraction}. The same design has been used to build the SOLARIS synchrotron light source in Krakow, Poland \cite{bartosik2013solaris}. Several other research labs in Brazil, France, Japan and United States have ongoing plans for building or upgrading storage rings based on a similar design \cite{reich2013ultimate}. The reduced emittance is expected to improve the spatial resolution below 10 nm, and the increased X-ray flux should enable more sensitive detection that, in the case of magnetism, will push the limits of XMCD microscopy towards the detection of magnetic moments from single ferromagnetic atoms. These capabilities may present completely new opportunities for the study of the novel spin physics at interfaces, as discussed above.

On the FELs side, there is great excitement about the capability of producing fully coherent and ultrafast X-ray photons. Lasing using the principle of self-amplified spontaneous emission (SASE) \cite{pellegrini19924,hogan1998measurements} was first demonstrated at FLASH in Hamburg in the XUV range \cite{ayvazyan2006first}. A few years later, groundbreaking SASE X-ray lasing in the soft and hard X-ray regime at the Linac Coherent Light Source (LCLS) at SLAC National Accelerator Laboratory and Stanford University in California, USA was demonstrated \cite{emma2010first}, which opened the way for the construction of other FELs around the world. SACLA at the Japan Synchrotron Radiation Research Institute (JASRI) is already operational \cite{ishikawa2012compact}, while XFEL in Hamburg, Germany \cite{altarelli2011european} and SwissFEL at the Paul Scherrer Institute (PSI) in Villigen, Switzerland \cite{patterson2010coherent} are expected to start operation in the coming months. The PAL-XFEL at the Pohang Accelerator Laboratory is being built in South Korea and it is expected to become available for users in 2017 \cite{kang2015status}. The FERMI Elettra free electron laser in Trieste, Italy \cite{allaria2010fermi} was the first facility to demonstrate full polarisation control of the XUV photon, which is of fundamental importance for the ultrafast imaging of magnetism. Polarisation control at soft X-rays was also recently demonstrated at the LCLS with the implementation of the Delta undulator \cite{higley2016femtosecond,lutman2016polarization}. LCLS-II \cite{galayda2014lcls}, the upgrade of LCLS to high-repetition rate operation, is expected to bring another quantum leap in the spectroscopy and imaging capabilities of X-ray lasers using lensless techniques.

\section{Conclusion}
In summary, we reviewed the latest development in magnetic X-ray microscopy aimed at understanding magnetism at the nanoscale. STXM at resonant absorption edges is a particularly suitable tool for this purpose because of its combined high magnetic sensitivity, spatial and temporal resolutions. It is also an instrument that is commercially available, is currently installed at most synchrotron facilities, and has a performance that is steadily improving. Another advantage is the relatively simple data analysis that is required to retrieve images, which greatly speeds up data collection.

We revised the latest experimental results that used time-resolved XMCD techniques combined with STXM to image the motion of spins at the nanoscale. We first discussed the detection of spin currents flowing across metallic nanostructures through the detection of spin transfer torque switching in ferromagnets and spin accumulation in non-magnetic metals. Then, we presented the latest developments in the capabilities of detecting high-frequency magnetisation dynamics, which have been used to make movies of magnetic vortex dynamics and of spin waves excited by spin transfer torque. In all of these experiments, the direct observation of the physics at play brought forth new understanding that was not accessible in other indirect measurement techniques.

Finally, we presented an overview of the future opportunities using the new X-ray sources that have recently been built or that are under construction: diffraction limited storage rings and free electron lasers. It is anticipated that the boost in average or peak X-ray brightness that are provided by these new sources will offer a whole new range of possibilities for the study of spin physics with ever increasing sensitivity, speed and spatial resolution. Ultimately, it is expected that these new capabilities will allow us to perform experiments with nanometre and femtosecond resolution. Hence, allowing us to look directly at the temporal and spatial length scales that are relevant to magnetism in condensed matter, with implications for both our fundamental understanding and our ability to build spin-based devices.

\ack
I am grateful to Peter Fischer, Matthias Kl\"aui, Jan Luning, Guido Meier, Gisela Sch\"utz, Hermann St{\"o}ll and Markus Weigand for the useful discussions.
Support from the Swedish Research Council grant E0635001 and the Marie Sklodowska Curie Actions, Cofund, Project INCA 600398s is gratefully acknowledged.

\end{document}